\documentclass[sigconf,authorversion,nonacm]{acmart}
\usepackage{graphicx} 
\usepackage{subcaption}
\makeatletter

\DeclareCaptionLabelFormat{mysublabelfmt}{(\alph{sub\@captype})}
\makeatother
\usepackage[capitalise, noabbrev]{cleveref}
\usepackage{float}
\usepackage[inline]{enumitem}
\usepackage[para]{footmisc}
\usepackage{tablefootnote}

\title{Empirical Analysis of Temporal and Spatial Fault Characteristics in Multi-Fault Bug Repositories}
\author{Dylan Callaghan}
\affiliation{%
  \institution{Stellenbosch University}
  \city{Stellenbosch}
  \country{South Africa}
}
\email{21831599@sun.ac.za}
\author{Alexandra van der Spuy}
\affiliation{%
  \institution{Stellenbosch University}
  \city{Stellenbosch}
  \country{South Africa}
}
\email{23552395@sun.ac.za}
\author{Bernd Fischer}
\affiliation{%
  \institution{Stellenbosch University}
  \city{Stellenbosch}
  \country{South Africa}
}
\email{bfischer@sun.ac.za}
\date{February 2025}


\begin{document}

\begin{abstract}
  Fixing software faults contributes significantly to the cost of
  software maintenance and evolution. Techniques for reducing these costs
  require datasets of software faults, as well as an understanding of the
  faults, for optimal testing and evaluation. In this paper, we present an empirical
  analysis of the temporal and spatial characteristics of faults existing in 16
  open-source Java and Python projects, which form part of the Defects4J and BugsInPy
  datasets, respectively. Our findings show that many faults in these software
  systems are long-lived, leading to the majority of software versions having
  multiple coexisting faults. This is in contrast to the assumptions of the
  original datasets, where the majority of versions only identify a single fault.
  In addition, we show that although the faults are found in only a small subset
  of the systems, these faults are often evenly distributed amongst this subset,
  leading to relatively few bug hotspots.
\end{abstract}

\maketitle

\section{Introduction}
Throughout the lifetime of a software system, a large number of faults
manifest themselves, where a fault (also called a bug or defect) is
defined as a deviation between the requirements and the actual system behavior.
The removal of faults forms a large part of the software
maintenance process~\cite{DBLP:journals/queue/ODell17}. A variety of factors
affect the time of discovery of a fault in the system~\cite{cost-drivers} and
the cost of software debugging (i.e., removing faults), and many faults remain
in the system for long periods of
time~\cite{Canfora_Ceccarelli_Cerulo_Di_Penta_2011}. In order to validate and understand
this phenomenon better, we conduct an empirical study using two datasets of software faults
mined from real-world open-source software systems, namely Defects4J \cite{defects4j}
and BugsInPy \cite{bugsinpy}. These datasets are commonly used in the validation
of software maintenance and debugging techniques. We build on work by Callaghan
and Fischer~\cite{datasets} and analyze these repositories based on the identification
of \emph{multiple faults} across the lifetime of the software, enabling an evaluation
of the faults throughout the evolution of the software system.

Our results confirm the findings in current literature~\cite{Canfora_Ceccarelli_Cerulo_Di_Penta_2011,
Saha_Khurshid_Perry_2015, Kim_Whitehead_2006, Marks_Zou_Hassan_2011} that a large
portion of faults are fixed within a short time after they are introduced, typically within a
day to a week.
However, the fault lifetimes exhibit a long-tail characteristic, and many faults
remain in the system for months or even years after their introduction.
The fault lifetimes roughly follow the expected negative exponential distribution; however,
we show that their logarithms have a zero-inflated, left-skewed distribution shape, which indicates that a more complex mathematical model may yield a better fit and thus allow better predictions.

We show that, correlated with (and presumably caused by) the large fault lifetimes, the majority
of system versions contain multiple simultaneous faults, with between 3 and
10 faults identified on average in each version. This is in contrast to the assumptions of the
original datasets, where the majority of versions only identify a single fault, but consistent with
the results shown by An et al.~\cite{multid4j} for the Java-based Defects4J benchmark suite.
%
Our analysis further shows that the number of faults that occur in a module over the system's lifetime is \emph{not} correlated with the module's size; instead, our findings indicate that the overall system faultiness can be attributed to the evolution of the system and, specifically, to different stages in the development workflow.

Finally, we show that only a small percentage (16--17\%) of the modules in each
project contain identified faults at all, tightening earlier observations about
the Pareto-principled nature of fault distributions (i.e., that 80\% of the faults
occur in only 20\% of the modules)~\cite{DBLP:conf/metrics/MollerP93, walkinshaw_are_2018,
Fenton_Ohlsson_2000, Andersson_Runeson_2007}. We also show that the number of faults
per module varies substantially per project: a large number of projects
have only a single fault per module, while other projects have one
or more modules with much higher fault occurrences.
%
We then give an intuitive definition of
\emph{bug hotspots} and show that they occur only rarely, with only 5 bug hotspots in 4 of the 16 projects.


To facilitate open science, we make our results publicly available in a full
replication package~\cite{replication}.


\section{Background}\label{sec:background}
In this study, we consider the open-source datasets Defects4J~\cite{defects4j}
and BugsInPy~\cite{bugsinpy}, which are real-world repositories of Java and
Python software projects with faults, respectively. We choose these datasets
firstly due to their popularity in the field of software debugging, as any findings
would directly affect this field, providing useful insights. We choose them secondly
due to their maturity and support, which enables a detailed analysis.
The datasets are comprised of bug entries, which refer to two versions in the
underlying project repository, a faulty version with a single fault and fixed
version, as well as the test suite exposing the fault. The entries for each project
are numbered from $1$ to $N$, where a lower number corresponds to a more recent
version pair.
\cref{tab:dataset} gives the overall dataset statistics for the projects in
Defects4j and BugsInPy. Note that we do not consider in our analysis projects with 10 or fewer versions from BugsInPy~\cite{bugsinpy},
due to their limiting size.

In order to aid in the analysis of this study, we build upon the work of An et
al.~\cite{multid4j} and Callaghan and Fischer~\cite{datasets}. An et al. identify
which Defects4J faults may be present in Defects4J versions other than the
initially identified version for that fault. They check for
\emph{exposure} of each fault (i.e., failure of at least one test case related to the fault) in each version,
and deem a fault to be present in a version if it is \emph{exposed}.
Callaghan and Fischer build on the work of An et al. by including the source code
locations of these faults in each version they are present, as well as extending
the technique to the BugsInPy dataset.

%


\begin{table}[!htb]
    \centering
    \caption{Dataset statistics. $N$ is the number of versions in the project;
      the number of faults and program sizes are averaged over all project versions.
    }\label{tab:dataset}
    {\footnotesize \tabcolsep2pt \begin{tabular}{|l|r|rrr|rrr|rrr|}\hline
         & & \multicolumn{3}{c|}{Modules} & \multicolumn{3}{c|}{Methods}  & \multicolumn{3}{c|}{KLOC}\\
        Project & $N$
        & Min & Mean & Max & Min & Mean & Max
        & Min & Mean & Max\\
        \hline
        Chart\tablefootnote{\url{https://www.jfree.org/jfreechart}} & 26
        & 553 & 573.2 & 633 & 7129 & 7582.6 & 8661 & 203.3 & 208.7  & 232.4\\
        Closure\tablefootnote{\url{https://developers.google.com/closure/compiler/}} & 106
        & 298 & 383.3 & 413 & 3704 & 5523.4 & 6229 & 99.4  & 208.4  & 269.2\\
        Lang\tablefootnote{\url{https://commons.apache.org/proper/commons-lang/}} & 65
        & 76 & 89.0 & 108 & 1857 & 2036.5 & 2217 & 48.0  & 53.2   & 61.1\\
        Math\tablefootnote{\url{https://commons.apache.org/proper/commons-math/}} & 106
        & 136 & 508.5 & 808 & 1138 & 4321.2 & 6603 & 30.5  & 121.7  & 185.3\\
        Time\tablefootnote{\url{https://www.joda.org/joda-time/}} & 27
        & 156 & 156.3 & 157 & 3573 & 3657.6 & 3744 & 70.2  & 78.0   & 99.2\\
        \hline
        \textbf{Total Java} & 330
        & 76 & 362.8 & 808 & 1138 & 4431.9 & 8661 & 30.5  & 134.0  & 269.2\\
        \hline\hline
        ansible\tablefootnote{\url{https://github.com/ansible/ansible}} & 18
        & 439 & 3702.1 & 5190 & 3287 & 28613.7 & 39898
        & 101.7 & 1124.7 & 1590.1\\
        black\tablefootnote{\url{https://github.com/psf/black}} & 23
        & 14 & 20.4 & 24 & 211 & 288.2 & 404
        & 5.2   & 66.5   & 96.0\\
        fastapi\tablefootnote{\url{https://github.com/tiangolo/fastapi}} & 16
        & 21 & 27.6 & 40 & 80 & 110.8 & 130
        & 2.8   & 4.2    & 5.0\\
        keras\tablefootnote{\url{https://keras.io}} & 45
        & 71 & 78.9 & 90 & 1533 & 1626.2 & 1782
        & 36.6  & 39.5   & 42.4\\
        luigi\tablefootnote{\url{https://github.com/spotify/luigi}} & 33
        & 51 & 75.5 & 99 & 988 & 1364.4 & 1936
        & 14.2 & 20.1   & 28.8\\
        matplotlib\tablefootnote{\url{https://matplotlib.org/}} & 30
        & 139 & 142.0 & 150 & 5502 & 5576.7 & 5693
        & 118.3 & 120.7  & 123.3\\
        pandas\tablefootnote{\url{https://github.com/pandas-dev/pandas}} & 169
        & 209 & 222.3 & 233 & 5456 & 5525.4 & 5673
        & 159.4 & 161.7  & 164.8\\
        scrapy\tablefootnote{\url{https://github.com/scrapy/scrapy}} & 40
        & 171 & 229.5 & 247 & 1279 & 1557.9 & 1677
        & 15.6  & 20.4   & 22.6\\
        thefuck\tablefootnote{\url{https://github.com/nvbn/thefuck}} & 32
        & 59 & 116.0 & 195 & 194 & 368.2 & 588
        & 1.6   & 3.7    & 6.2\\
        tornado\tablefootnote{\url{https://github.com/tornadoweb/tornado}} & 16
        & 78 & 80.5 & 84 & 2818 & 3041.3 & 3423
        & 21.2  & 23.0   & 24.4\\
        youtube-dl\tablefootnote{\url{https://github.com/ytdl-org/youtube-dl}} & 43
        & 186 & 608.4 & 815 & 651 & 1936.0 & 2868
        & 20.5  & 82.6   & 138.0\\
        \hline
        \textbf{Total Python} & 465
        & 14 & 340.9 & 5190 & 80 & 4271.6 & 39898
        & 1.6    & 104.7  & 1590.1\\
        \hline
    \end{tabular}}
\end{table}

\section{Related Work}\label{sec:relatedwork}
There is a large body of work addressing the characteristics of faults in
software~\cite{DBLP:journals/ese/TanLLWZZ14}. However, we limit our study to the
analysis of temporal and spatial characteristics of faults.
Chou et al.~\cite{DBLP:conf/sosp/ChouYCHE01} provide an in-depth analysis on
operating systems of most spatial and temporal characteristics investigated
in this paper, and we largely corroborate their findings. In particular, they find
a substantial and increasing number of faults are available in each version
over time. Additionally, they find that some faults were fixed in a short time, while
others lasted much longer. Lastly, they find that faults are found in a small percentage
of the code, with a few files having a large number of faults, and many files having
one or two errors.

Related to the lifespan or time between introduction and fix of a fault, Saha et
al.~\cite{Saha_Khurshid_Perry_2015} find that there are a significant number of
long-lived faults (faults surviving more than a year), although most faults are fixed
within a day or week of their introduction. Marks et al.~\cite{Marks_Zou_Hassan_2011}
consider the lifespan of faults in the Mozilla and Eclipse projects and find
that the majority of faults are fixed within three months, after which faults
survive up to a few years. Canfora et al.~\cite{Canfora_Ceccarelli_Cerulo_Di_Penta_2011}
also find similar results for four open source projects including Mozilla, OpenLDAP,
Eclipse and Vuze. Lastly, Kim and Whitehead~\cite{Kim_Whitehead_2006} consider bug
fix time for a number of versions in the ArgoUML and postgreSQL projects.

Many studies have investigated the spatial distribution of faults in software
systems. M{\"{o}}ller and Paulish \cite{DBLP:conf/metrics/MollerP93} find that a small subset of the code accounts
for the majority of the faults. This is confirmed by Walkinshaw and
Minku~\cite{walkinshaw_are_2018}, who conclude that the Pareto principle
does apply to fault distribution (i.e., roughly 80\% of the faults can be found
in 20\% of the code). Fenton and Ohlsson~\cite{Fenton_Ohlsson_2000}
concur that the majority of faults can be found in a small subset of the system,
but that despite this, it is not as a result of faulty modules being larger.
These results are confirmed by Andersson and Runeson~\cite{Andersson_Runeson_2007}
in their replication study using different systems. Lastly, Grbac and
Huljenic~\cite{GalinacGrbac_Huljenic_2015} show that the double Pareto distribution
is the best choice for spatial fault distribution.


\section{Bug Prevalence and Evolution}\label{sec:bpv}
In this section, we discuss the prevalence of multiple faults in the systems
under analysis, and seek to understand this by analyzing changes in the
number of faults over the evolution of the systems.

\subsection{Bug Prevalence}\label{sec:bug-prevalence}

\begin{figure*}[!htb]
    \centering
    \begin{subfigure}{0.24\textwidth}
        \centering
        \includegraphics[height=3cm]{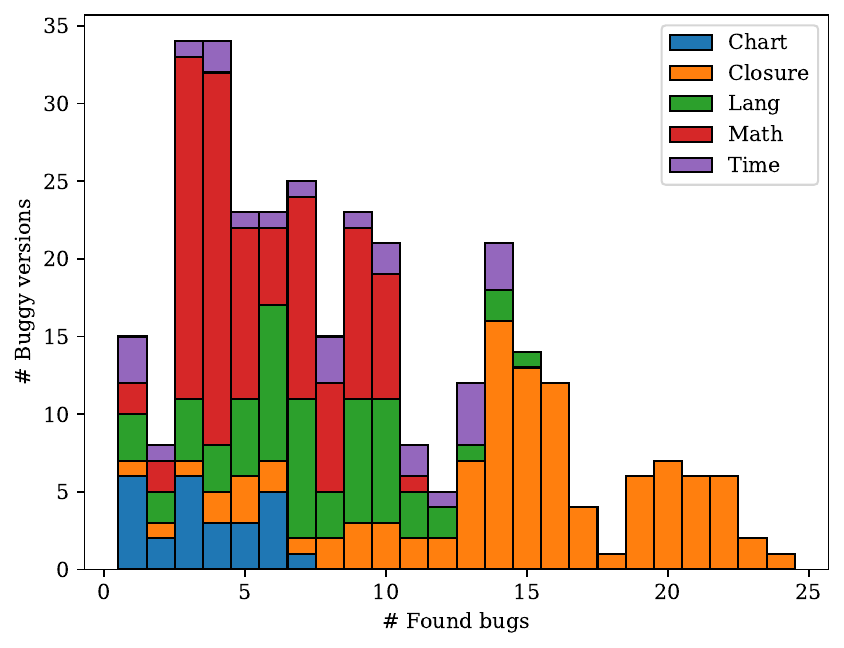}
        \caption{Defects4J}\label{fig:foundbugs:d4j_all}
    \end{subfigure}
    \hfil
    \begin{subfigure}{0.24\textwidth}
        \centering
        \includegraphics[height=3cm]{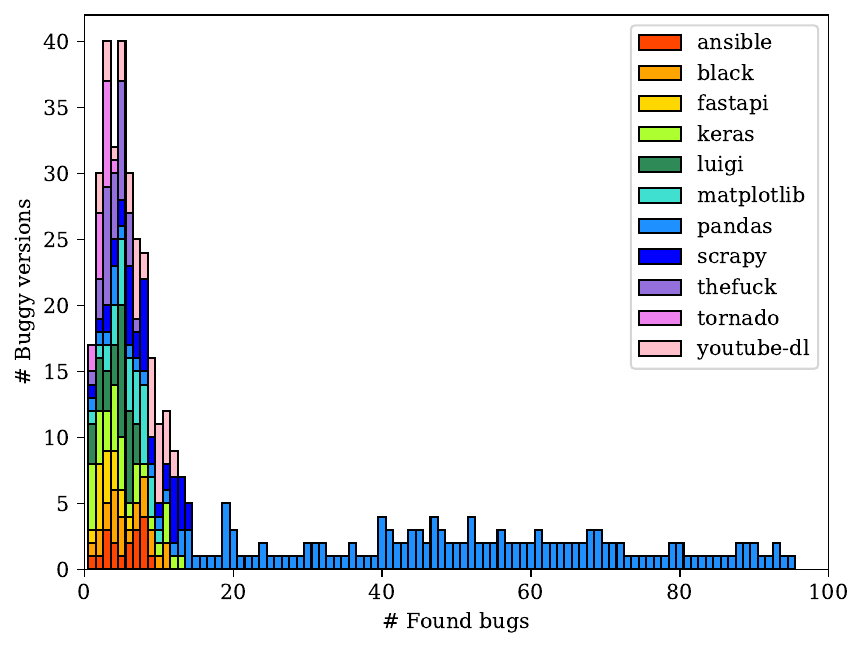}
        \caption{BugsInPy}\label{fig:foundbugs:bip_all}
    \end{subfigure}
    \hfil
    \begin{subfigure}{0.24\textwidth}
        \centering
        \includegraphics[height=3cm]{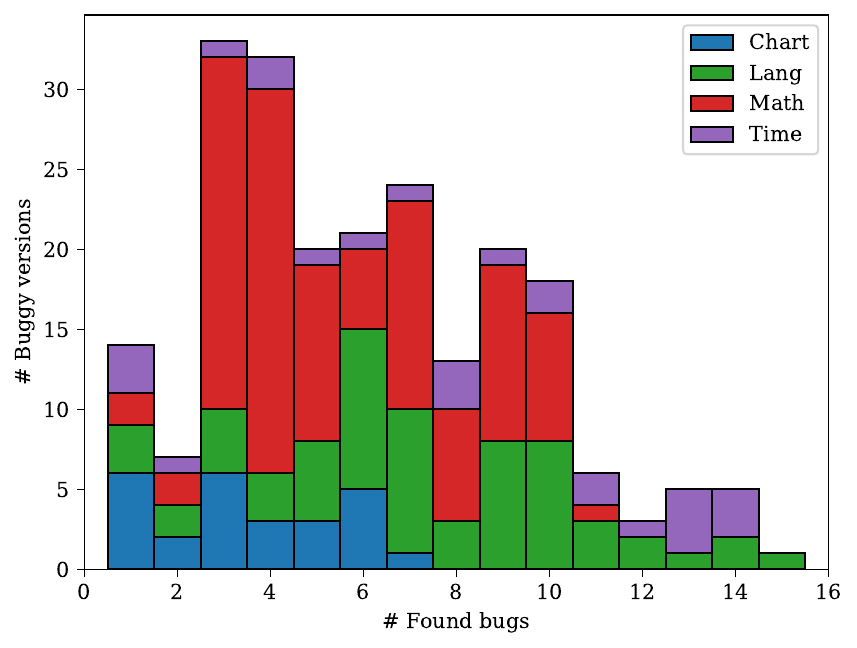}
        \caption{Defects4J (without \texttt{Closure})}\label{fig:foundbugs:d4j_out}
    \end{subfigure}
    \hfil
    \begin{subfigure}{0.24\textwidth}
        \centering
        \includegraphics[height=3cm]{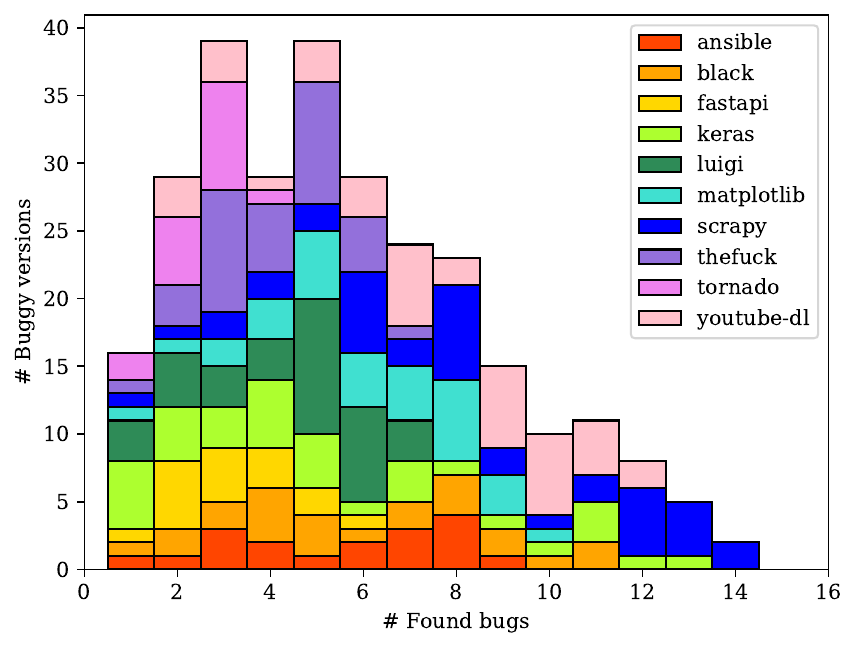}
        \caption{BugsInPy (without \texttt{pandas})}\label{fig:foundbugs:bip_out}
    \end{subfigure}
    \caption{The number of faulty versions in the multifault datasets over the number of faults}\label{fig:foundbugs}
\end{figure*}

\cref{fig:foundbugs} shows the distribution of the number of faults in each
version, given for each software project.
The first observation from \cref{fig:foundbugs} is that the majority of versions
contain multiple faults, with only 15 single-fault versions over all 330 Defects4J
versions; and 17 single-fault versions over all 498 BugsInPy versions. This
shows that although multiple faults are not often \emph{fixed} in the same
version, as noted by Perez et al.~\cite{Perez_Abreu_dAmorim_2017}, the majority
of versions do have multiple faults \emph{co-existing}.

We note that the distributions of both Defects4J and BugsInPy are each skewed by
a large project (namely \texttt{Closure} and \texttt{pandas} respectively) with many faulty
versions. \cref{fig:foundbugs:d4j_out,fig:foundbugs:bip_out} thus show these
distributions without these outliers in order to see more detail. From these
figures, we can see that the distribution of faults is roughly the same in all
projects from both Defects4J and BugsInPy, and that the differences between Defects4J
and BugsInPy (without outliers) are small: both distributions have a skew to lower
number of faults per version, with the majority of versions containing 3--10 faults.
However, we see that the outliers (\texttt{Closure} from Defects4J and \texttt{pandas} from
BugsInPy) have versions with a significantly higher number of faults and more even
distribution, with averages of 14.9 and 48.5 for Closure and pandas
respectively.
Correlating these results with the details for these projects presented in
\cref{tab:dataset}, we find that they are the largest in terms of size and
number of versions. We leave the discussion as
to why these projects may be outliers to \cref{sec:bug-evolution}.

\subsection{Bug Evolution}\label{sec:bug-evolution}

\begin{figure}[ht]
  \hspace{-0.06\columnwidth}
  \begin{subfigure}{0.4\columnwidth}
        \centering
        \includegraphics[height=3cm]{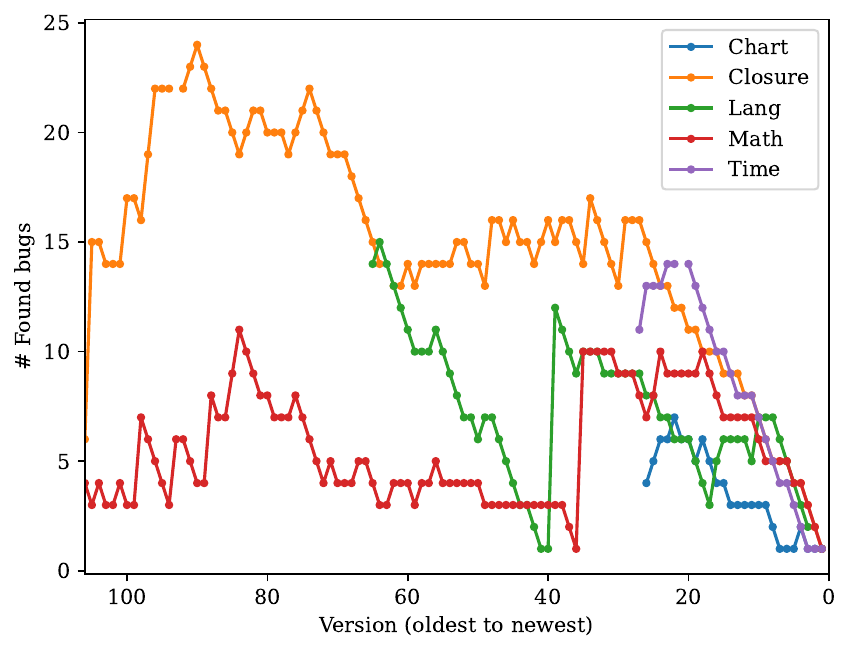}
        \caption{Defects4J}
    \end{subfigure}
    \hfil
    \begin{subfigure}{0.51\columnwidth}
        \centering
        \includegraphics[height=3cm]{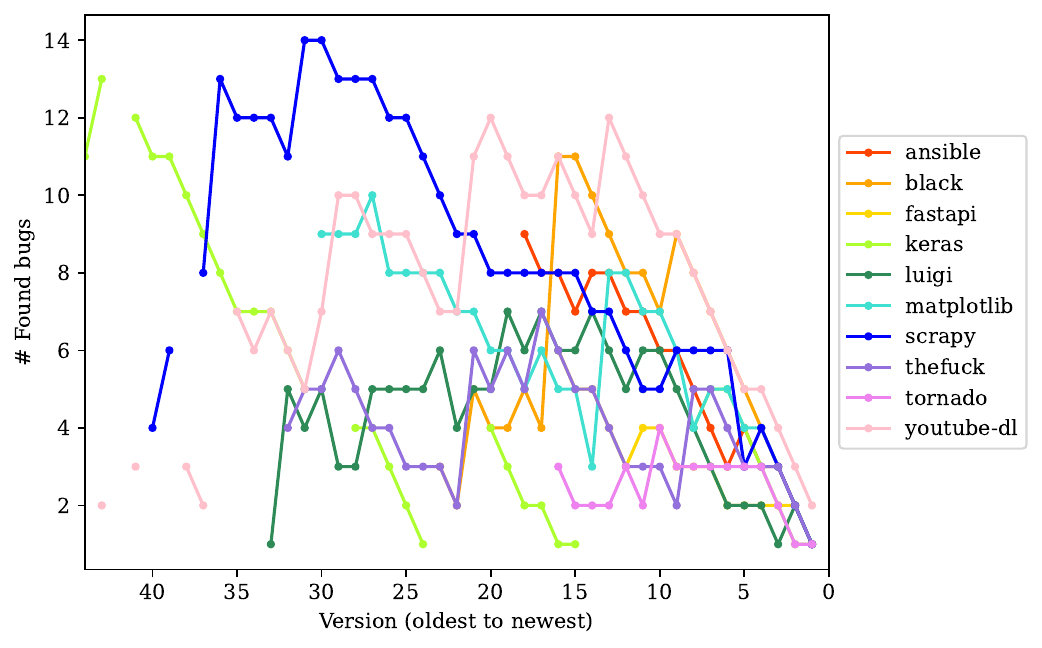}
        \caption{BugsInPy (without \texttt{pandas})}
    \end{subfigure}
    \caption{The number of identified faults in a version over all project versions, from oldest to newest per project. Discontinuities represent deprecated faults (Defects4J) or non-reproducible faults (BugsInPy)}\label{fig:versionbugs}
\end{figure}

In order to understand the causes behind versions with a large number of faults,
we analyze  in this
section the number of faults over the evolution of the system.
\cref{fig:versionbugs} shows the number of identified faults in each version over
the course of each project's history.
Although each project exhibits a unique history and evolution of the number of
faults, we see that all projects show a combination of three trends:
\begin{enumerate*}
  \item large increases in the number of faults between few versions (e.g., between
    \texttt{Lang} 40 and 41),
  \item gradual decreases between many versions (e.g., \texttt{youtube-dl} 11 to 5), and
  \item the number of faults remaining constant (e.g., \texttt{Math} 50 to 42)
\end{enumerate*}.
In order to understand
these three trends,
we analyzed the changes between each version in the
underlying Git repositories of the projects.

\begin{figure}[ht]
    \centering
    \begin{subfigure}{0.48\columnwidth}
        \centering
        \includegraphics[height=3cm]{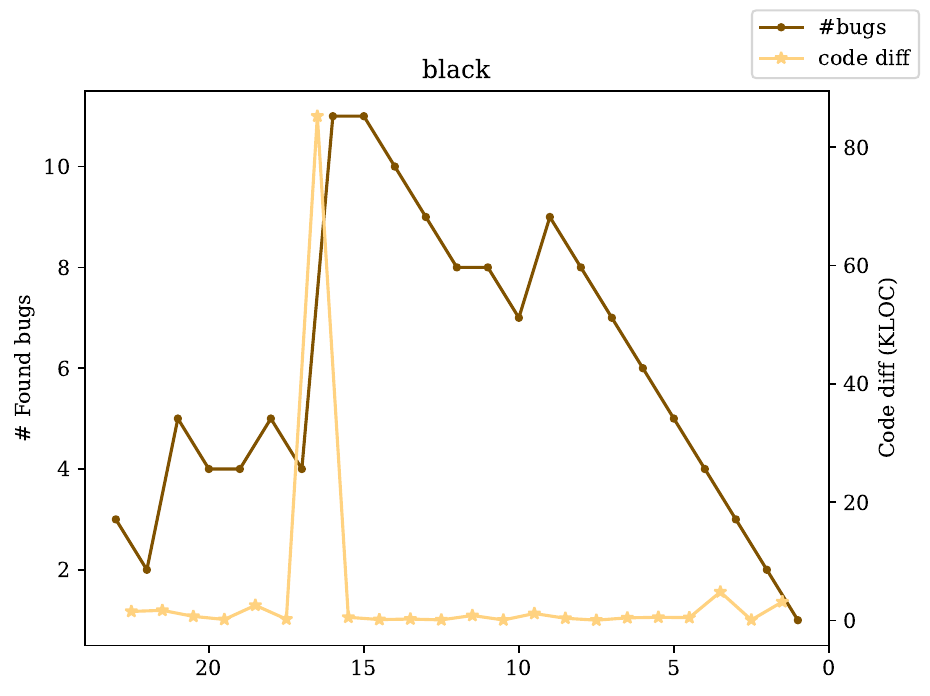}
        \caption{black}\label{fig:exampleversionbugs:black}
    \end{subfigure}
    \hfil
    \begin{subfigure}{0.48\columnwidth}
        \centering
        \includegraphics[height=3cm]{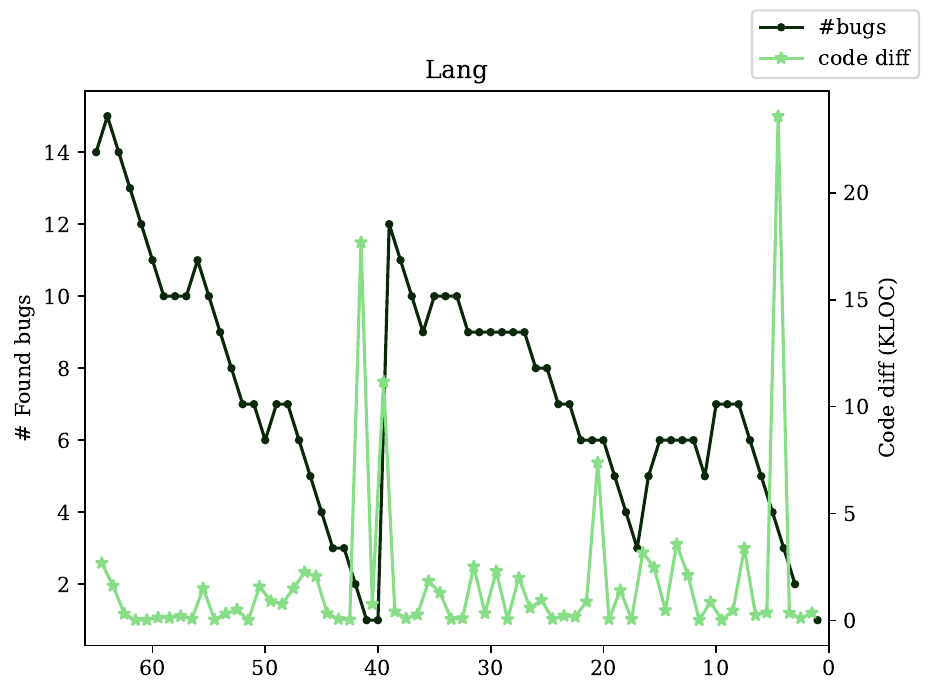}
        \caption{Lang}\label{fig:exampleversionbugs:Lang}
    \end{subfigure}
    \begin{subfigure}{0.48\columnwidth}
        \centering
        \includegraphics[height=3cm]{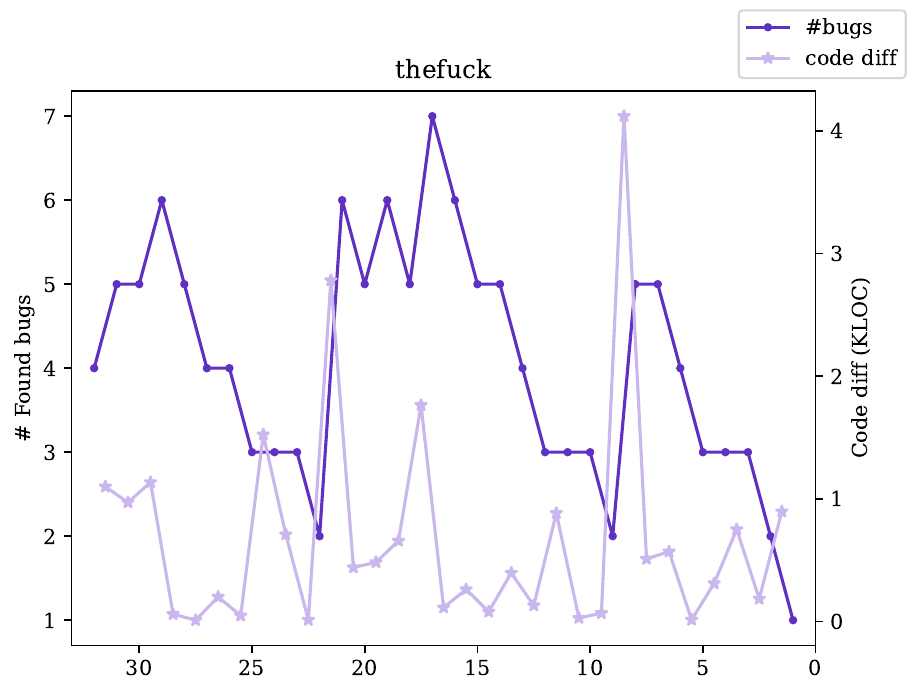}
        \caption{thefuck}\label{fig:exampleversionbugs:thefuck}
    \end{subfigure}
    \hfil
    \begin{subfigure}{0.48\columnwidth}
        \centering
        \includegraphics[height=3cm]{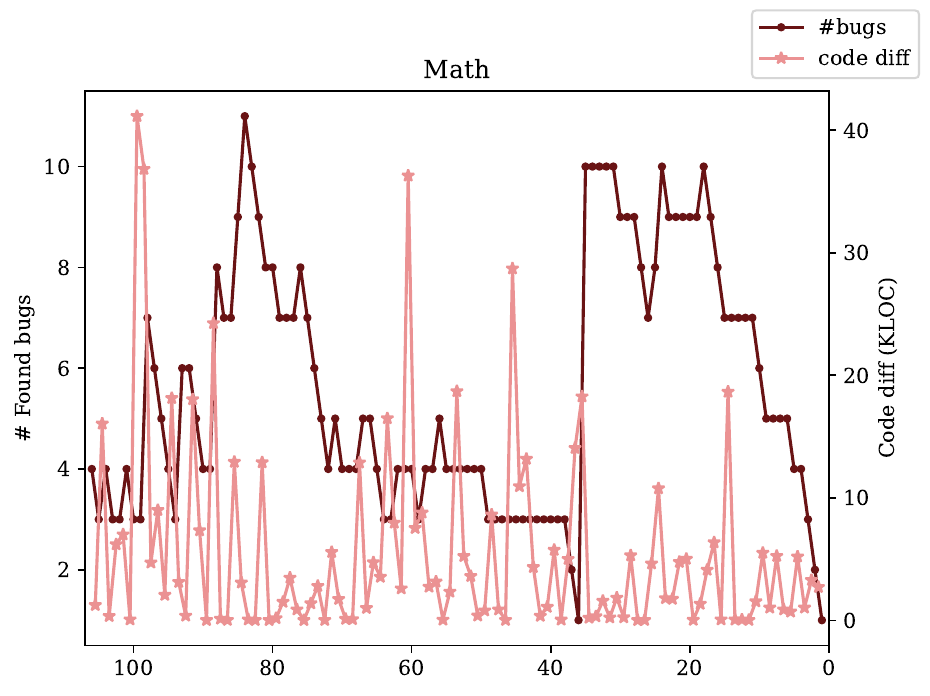}
        \caption{Math}\label{fig:exampleversionbugs:Math}
    \end{subfigure}
    \begin{subfigure}{0.48\columnwidth}
        \centering
        \includegraphics[height=3cm]{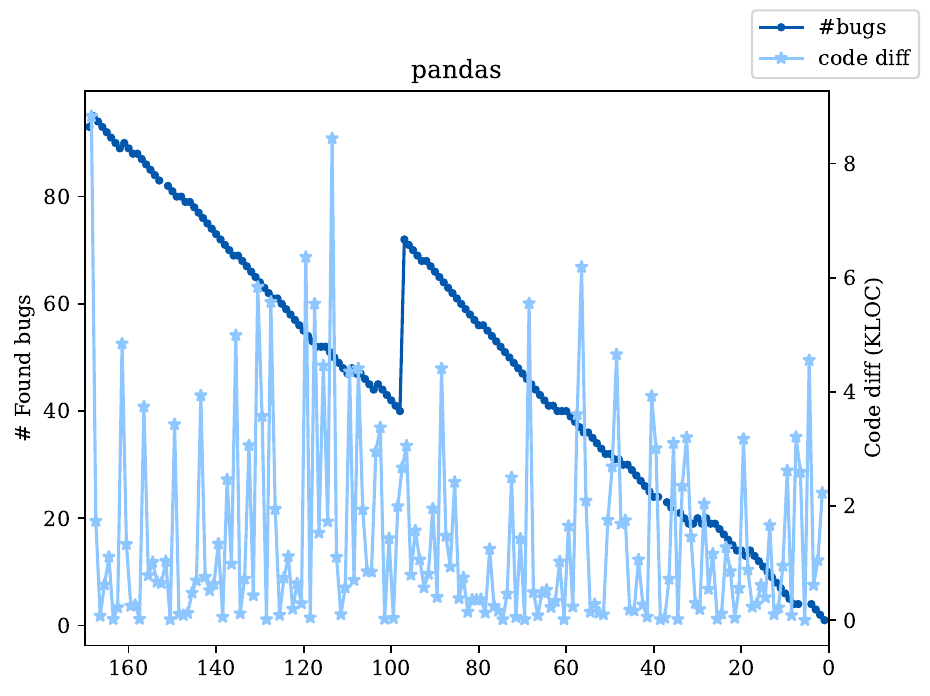}
        \caption{pandas}\label{fig:exampleversionbugs:pandas}
    \end{subfigure}
    \hfil
    \begin{subfigure}{0.48\columnwidth}
        \centering
        \includegraphics[height=3cm]{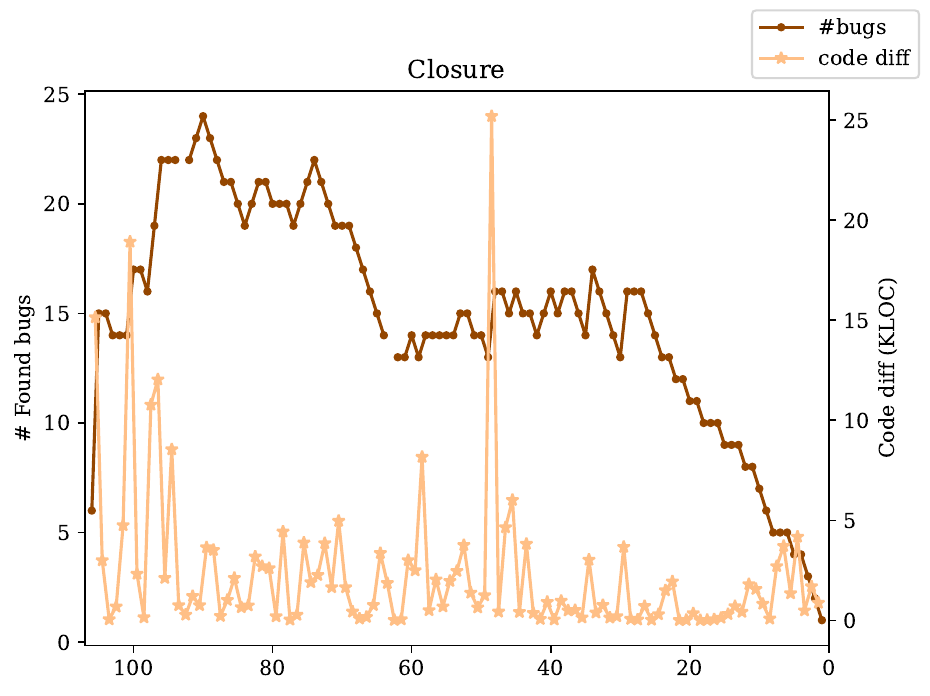}
        \caption{Closure}\label{fig:exampleversionbugs:Closure}
    \end{subfigure}
    \caption{Number of faults per version and code diff size over all
    versions (oldest to newest) for a subset of projects from BugsInPy (left) and
    Defecst4J (right)}\label{fig:exampleversionbugs}
\end{figure}

The plots in \cref{fig:exampleversionbugs} show the number of identified faults in
each version as well as the size of the diff between every two
versions, for a representative subset of the projects. For
\texttt{black} (cf.\ \cref{fig:exampleversionbugs:black}), we see that the spike in the number
of identified faults between versions 17 and 16 is accompanied by a large number of
code changes visible in the size of the diff between the two versions. In
contrast, the size of the diffs between any other two versions are much smaller,
which is during the time when the number of faults is being reduced (i.e., bug
fixing). A similar trend can be seen for \texttt{Lang} in \cref{fig:exampleversionbugs:Lang},
with the large spike between versions 41 and 39.
%
However, large changes in the system do not always signify fault additions, as visible
in the changes between versions 5 and 4.

The other projects (see \cref{fig:exampleversionbugs:thefuck,%
fig:exampleversionbugs:pandas,fig:exampleversionbugs:Math,fig:exampleversionbugs:Closure})
show a more representative but less obvious trend; large spikes or consistencies
in the number of faults is often (but not always) accompanied by larger changes
in the system (e.g., between versions 9 and 8 in \texttt{thefuck}), and steady decreases in
the number of faults often (but not always) coincide with small changes that indicate
bug-fixes.

We therefore conclude that large changes in the code lead to the introduction of
a large number of faults which are often fixed one-at-a-time, leaving some faults
to be fixed many versions later. Additionally, larger projects such as \texttt{Closure}
and \texttt{pandas} (see \cref{fig:exampleversionbugs:Closure,fig:exampleversionbugs:pandas})
tend to accumulate much more technical debt over time due to parts of the system
that remain unchanged, leading to a larger number of faults in their versions.
This gives an explanation for these projects being outliers in
\cref{fig:foundbugs:d4j_all,fig:foundbugs:bip_all} in \cref{sec:bug-prevalence}, as these are large
projects that have a large number of versions spanning a long version history.

\section{Bug Lifetimes}\label{sec:lifetimes}

\begin{figure*}[ht]
    \centering
    \begin{subfigure}{0.24\textwidth}
        \centering
        \includegraphics[height=3.4cm]{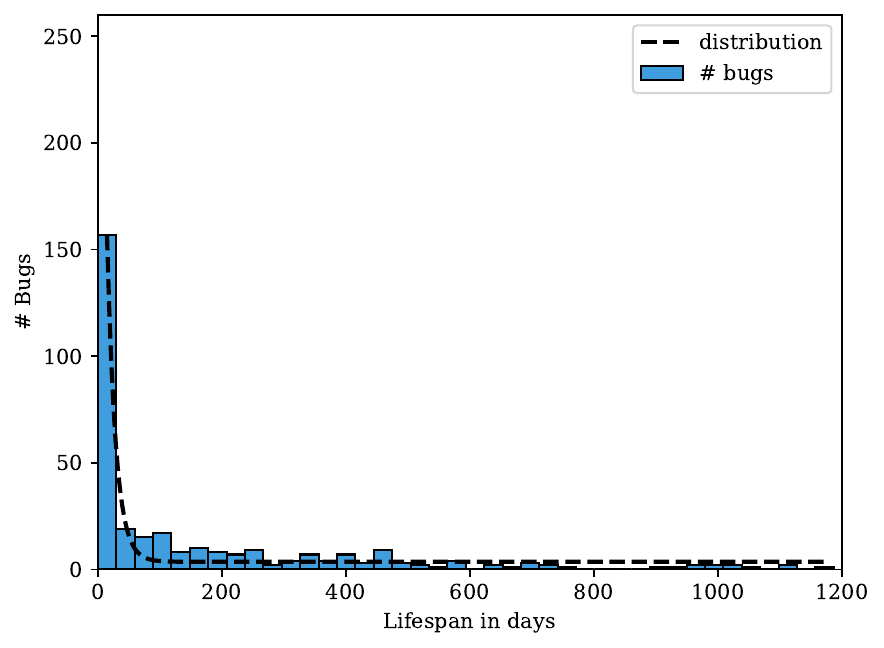}
      \caption{Defects4J}\label{fig:lifespan:d4je} 
    \end{subfigure}
    \hfil
    \begin{subfigure}{0.24\textwidth}
        \centering
        \includegraphics[height=3.4cm]{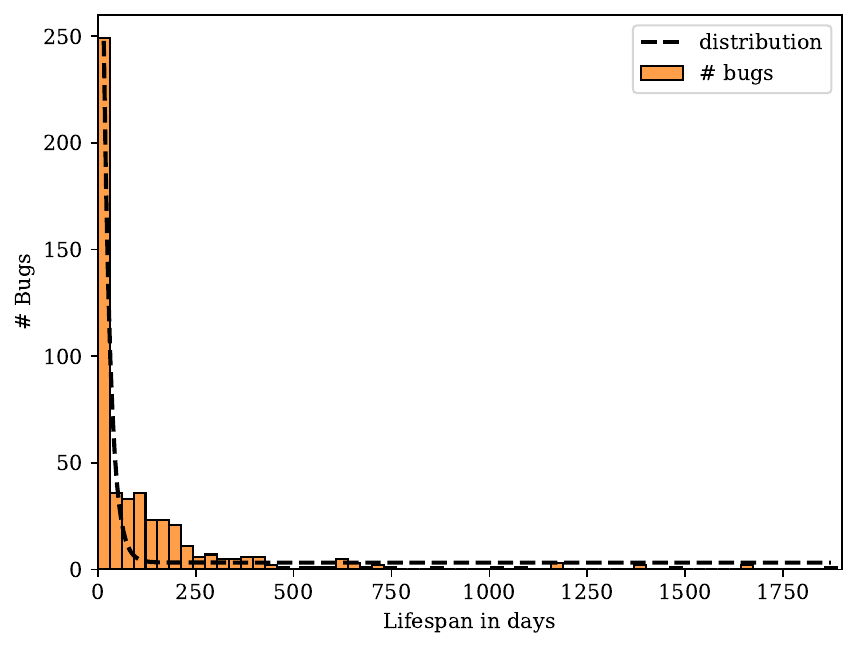}
        \caption{BugsInPy}\label{fig:lifespan:bipe} 
    \end{subfigure}
    \hfil
    \begin{subfigure}{0.24\textwidth}
        \centering
        \includegraphics[height=3.3cm]{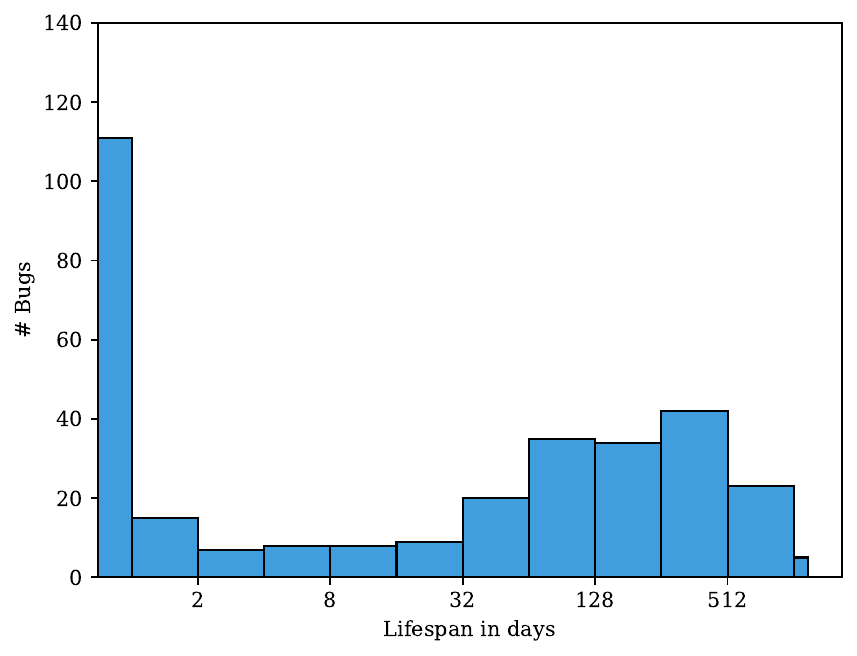}
        \caption{Defects4J ($\log_2$-scale)}\label{fig:lifespan:d4j}
    \end{subfigure}
    \hfil
    \begin{subfigure}{0.24\textwidth}
        \centering
        \includegraphics[height=3.3cm]{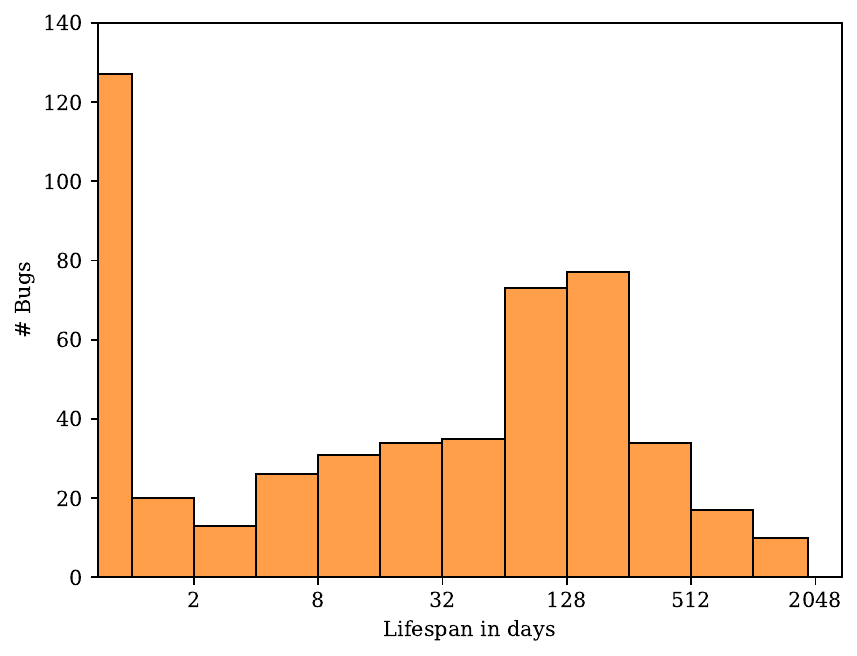}
        \caption{BugsInPy ($\log_2$-scale)}\label{fig:lifespan:bip}
    \end{subfigure}
    \caption{Fault lifetimes in Defects4J and BugsInPy}\label{fig:lifespan}
\end{figure*}

We define the lifetime of a fault as the amount of time between its 
introduction into the system and its fix. However, for the projects in our
analysis we do not know exactly when each fault was actually introduced. We
instead \emph{estimate} this by using the date of the last version it can successfully
be reproduced in. Note that this systematically underestimates the actual fault
lifetimes.

Although there is no general consensus in the related work
(cf.\ \Cref{sec:relatedwork}) about the average lifetime of a fault in a system,
there is a general trend that is clear from all results: A large number of faults
are fixed on the same day or within the same week, after which the majority of faults
tend to last much longer in the system, up to a few months or years.
We attribute the differences in results to the differences in development workflow
between the subject programs considered in each study.

\cref{fig:lifespan:d4je,fig:lifespan:bipe} show the lifetime in days for each fault
in our subject projects, binned into periods of roughly one month.
Our results conform to the general consensus observed from other
studies, i.e., that a large number of faults are fixed quickly, with
the lifetimes roughly following a long-tailed negative exponential distribution. The overlaid best-fit curve demonstrates this. This confirms the results of previous studies for the popular
fault datasets Defects4J and BugsInPy, which provides evidence of
the representativity of both Defects4J and BugsInPy of real-world faults, and
strengthens their claims of being realistic fault datasets.

However, a closer inspection of the fitted curve shows that it underestimates the number of faults in the long tail. To provide more detail here, we show the
lifetimes over a log-scaled x-axis in \cref{fig:lifespan:d4j,fig:lifespan:bip}, with logarithmic bins.
We clearly see the spike of same-day fixes, while lifetimes greater than a single day
have modes at the ranges of 256 to 512 resp.\ 128 to 256 days, with much smaller medians and means.
The logarithmic lifetimes thus have a zero-inflated, left-skewed distribution shape, which indicates that a more complex mathematical model may yield a better fit and thus allow better predictions.

A comparison between Defects4J in \cref{fig:lifespan:d4j}, and
BugsInPy in \cref{fig:lifespan:bip} shows that the bug lifetimes of both
datasets follow largely the same distribution. We thus conclude that the impact
of the programming language (Java or Python) is small, with the deviations between
projects of different languages more likely to be characteristic of the development
process than the language itself.


\section{Bug Hotspots and Density}\label{sec:hotspots}
\begin{figure*}[!htb]
    \centering
    \begin{subfigure}{0.32\textwidth}
        \centering
        \includegraphics[height=3cm]{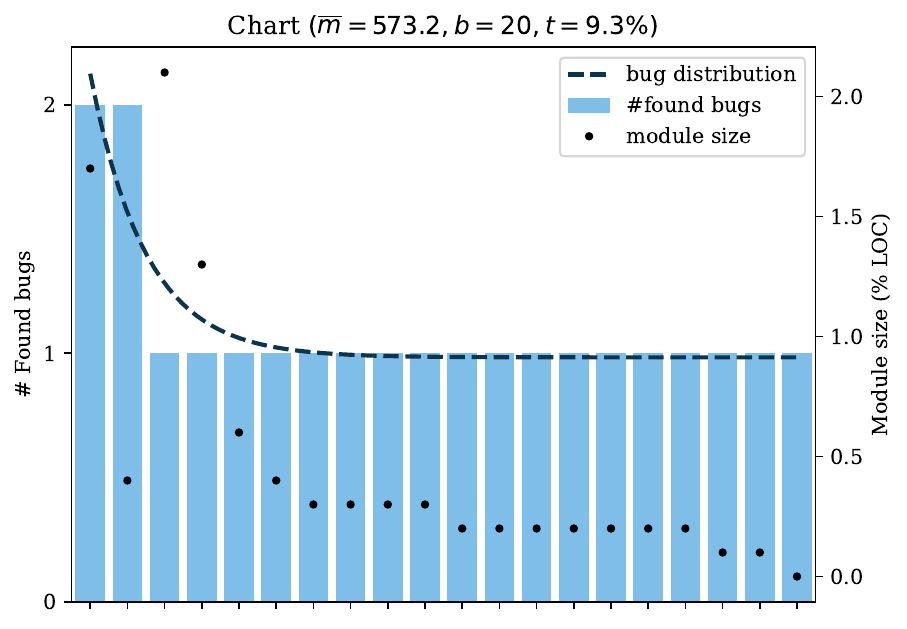}
    \end{subfigure}
    \begin{subfigure}{0.32\textwidth}
        \centering
        \includegraphics[height=3cm]{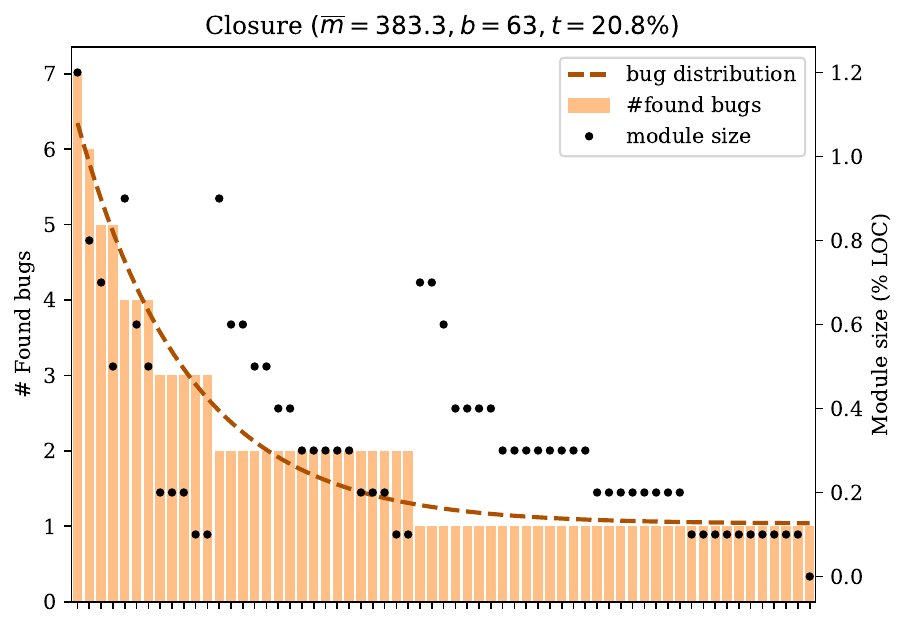}
    \end{subfigure}
    \begin{subfigure}{0.32\textwidth}
        \centering
        \includegraphics[height=3cm]{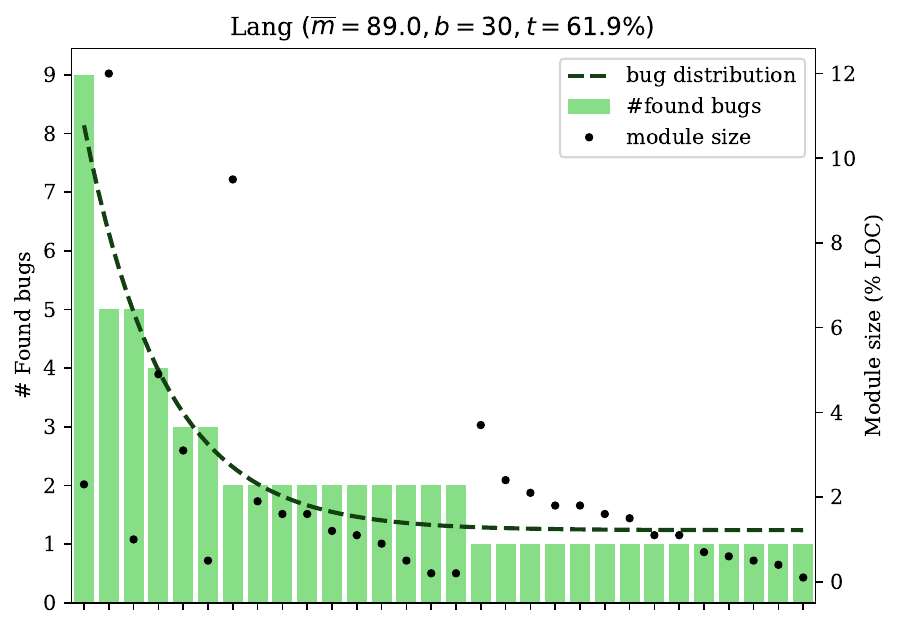}
    \end{subfigure}
    \begin{subfigure}{0.32\textwidth}
        \centering
        \includegraphics[height=3cm]{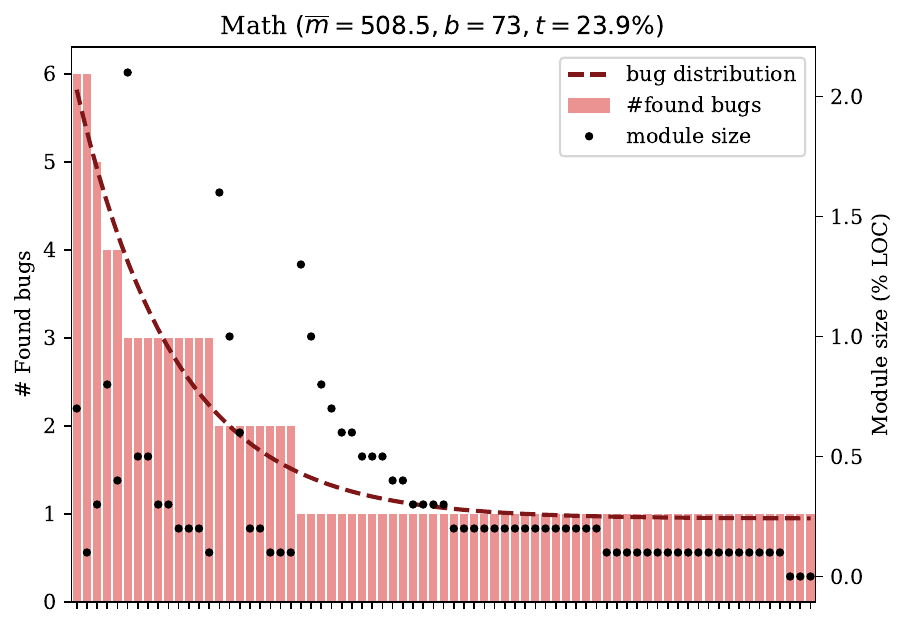}
    \end{subfigure}
    \begin{subfigure}{0.32\textwidth}
        \centering
        \includegraphics[height=3cm]{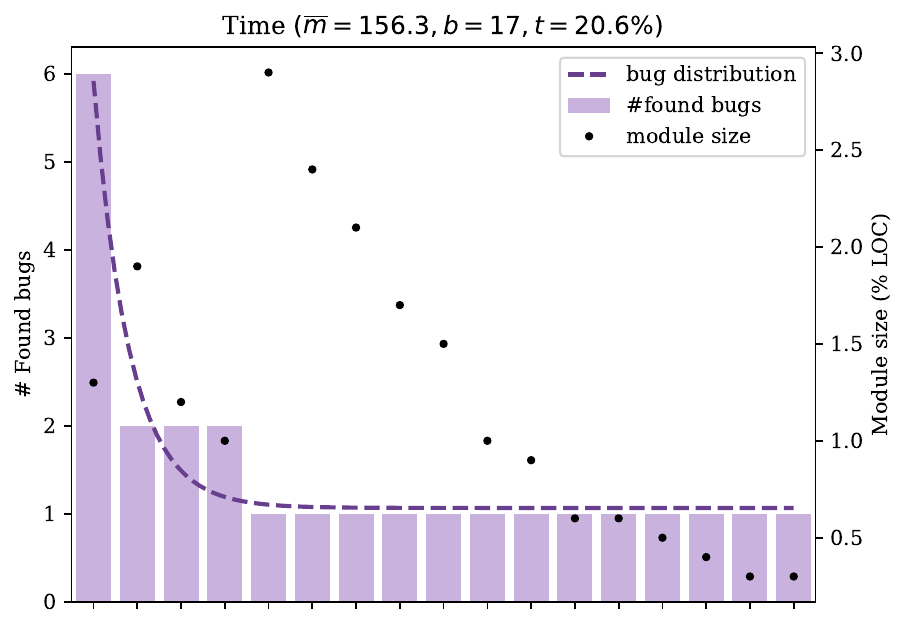}
    \end{subfigure}
    \caption{The number of identified faults in each faulty module for Defects4J
      projects sorted by decreasing number of faults then size
      ($\overline{m}$: mean total number of modules, \textit{b}: number of faulty modules,
      \textit{t}: total faulty modules size (\% LOC))}\label{fig:d4jmodule}
\end{figure*}

\begin{figure*}[!htb]
    \centering
    \begin{subfigure}{0.24\textwidth}
        \centering
        \includegraphics[height=3cm]{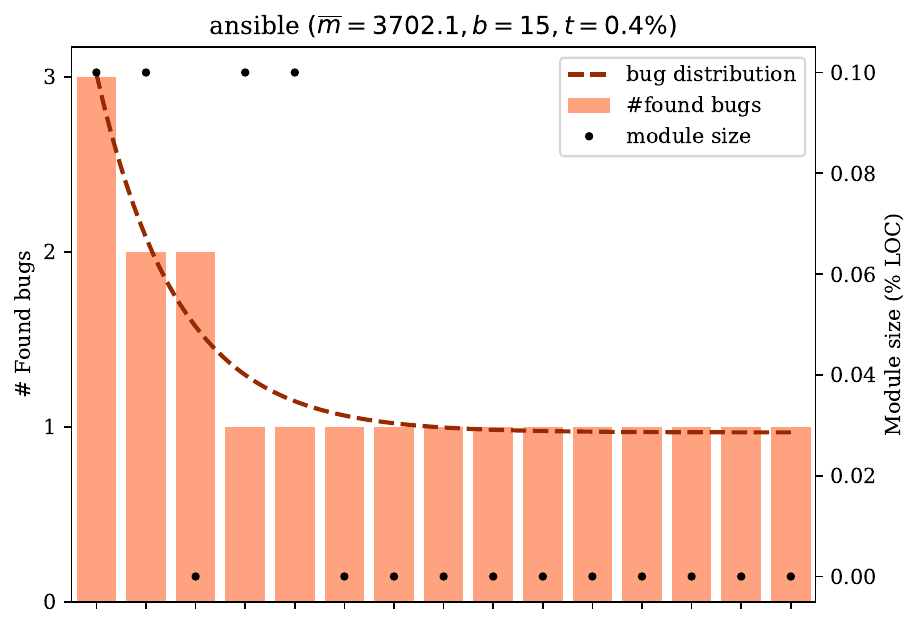}
    \end{subfigure}
    \begin{subfigure}{0.24\textwidth}
        \centering
        \includegraphics[height=3cm]{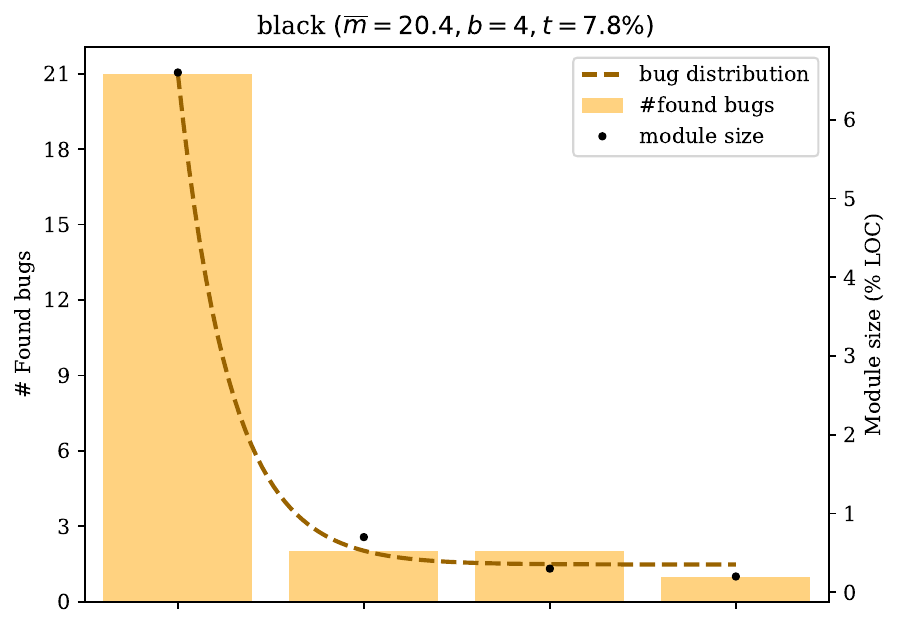}
    \end{subfigure}
    \begin{subfigure}{0.24\textwidth}
        \centering
        \includegraphics[height=3cm]{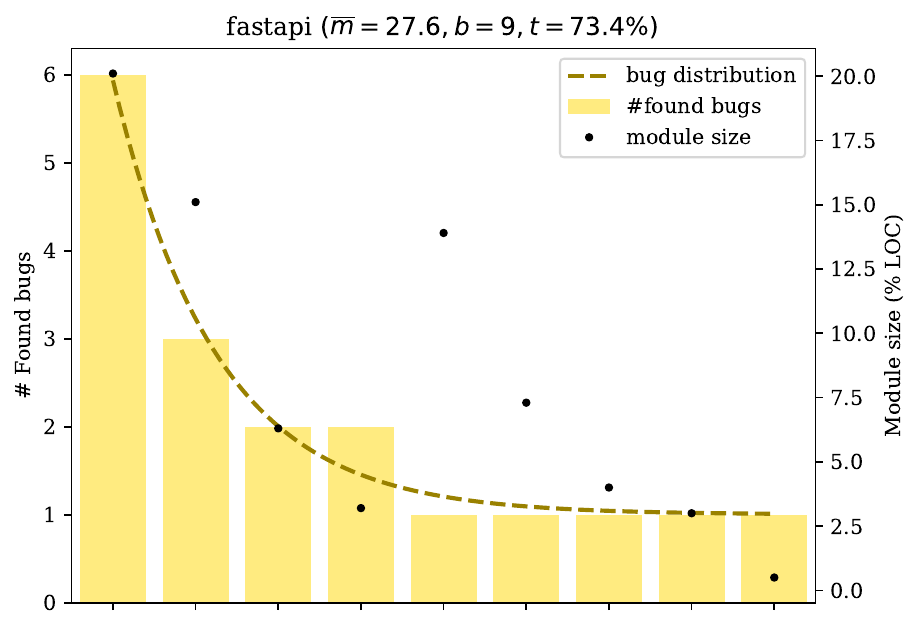}
    \end{subfigure}
    \begin{subfigure}{0.24\textwidth}
        \centering
        \includegraphics[height=3cm]{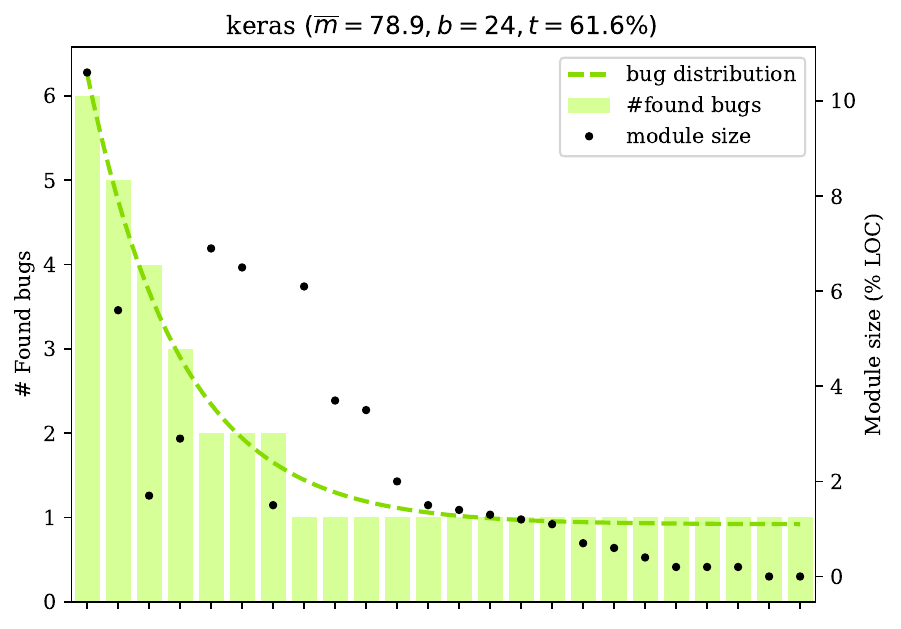}
    \end{subfigure}
    \begin{subfigure}{0.24\textwidth}
        \centering
        \includegraphics[height=3cm]{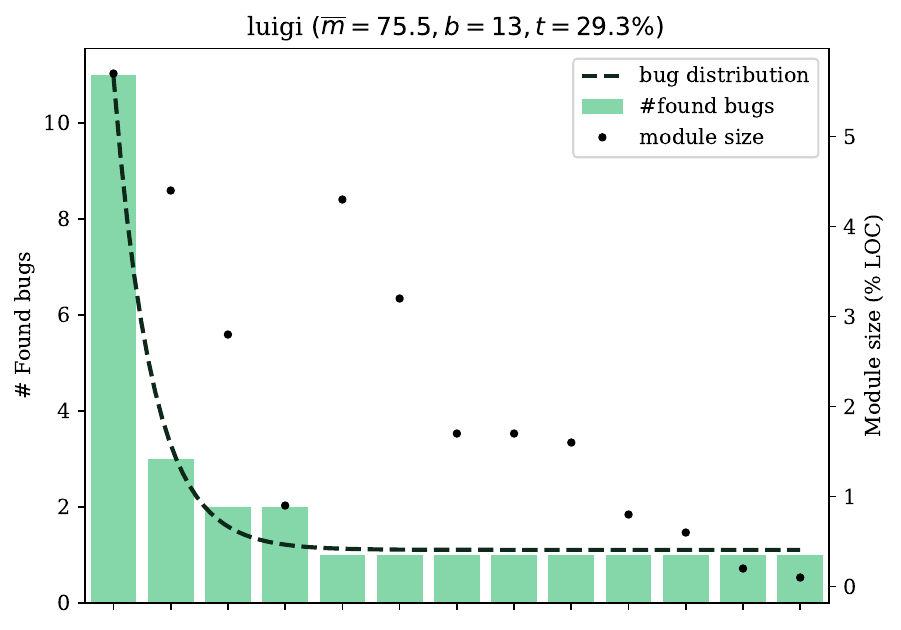}
    \end{subfigure}
    \begin{subfigure}{0.24\textwidth}
        \centering
        \includegraphics[height=3cm]{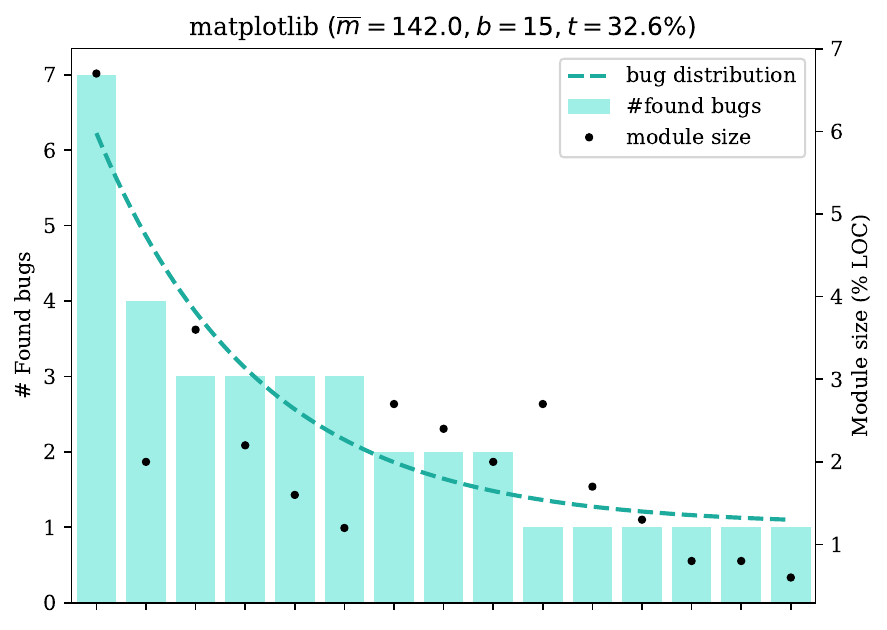}
    \end{subfigure}
    \begin{subfigure}{0.24\textwidth}
        \centering
        \includegraphics[height=3cm]{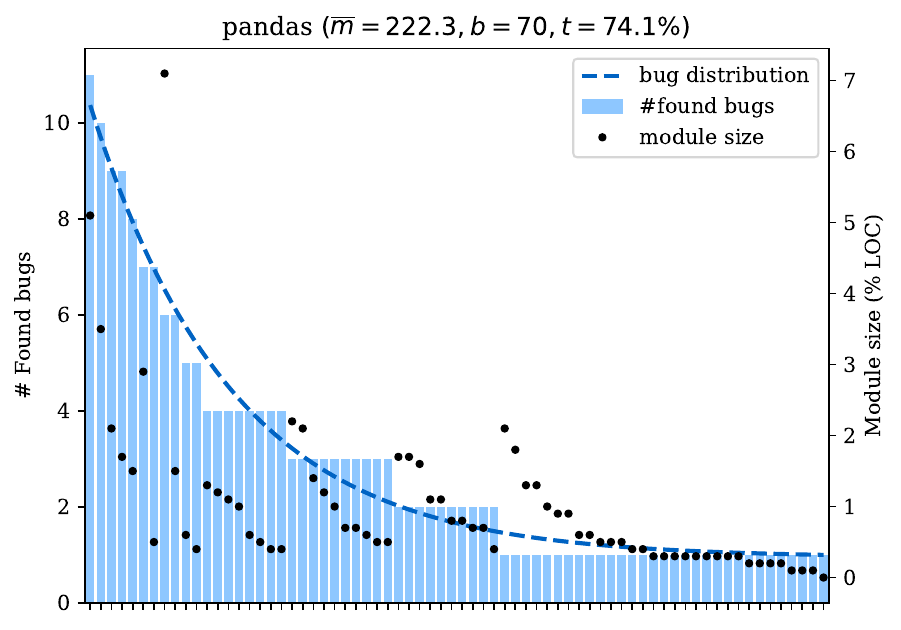}
    \end{subfigure}
    \begin{subfigure}{0.24\textwidth}
        \centering
        \includegraphics[height=3cm]{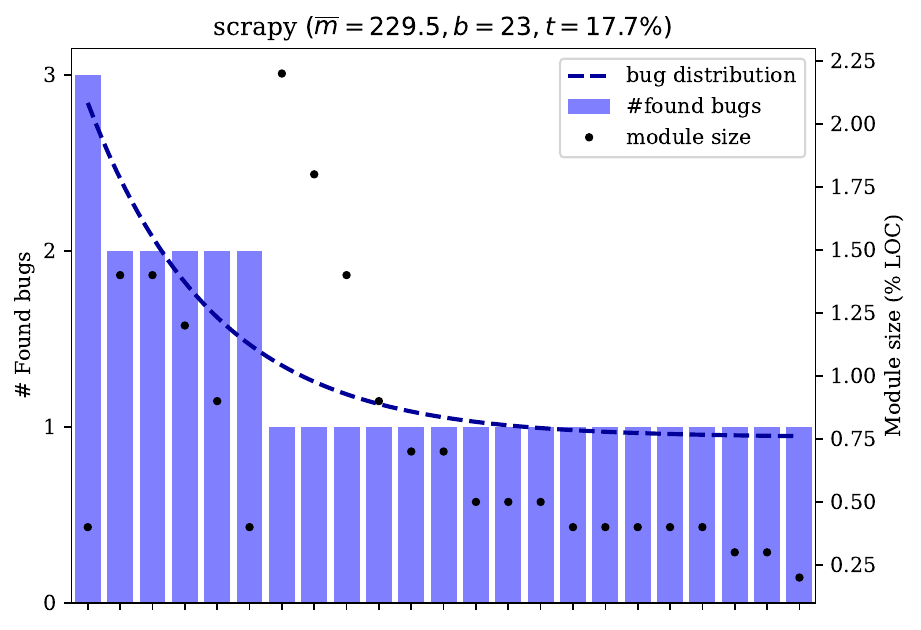}
    \end{subfigure}
    \begin{subfigure}{0.24\textwidth}
        \centering
        \includegraphics[height=3cm]{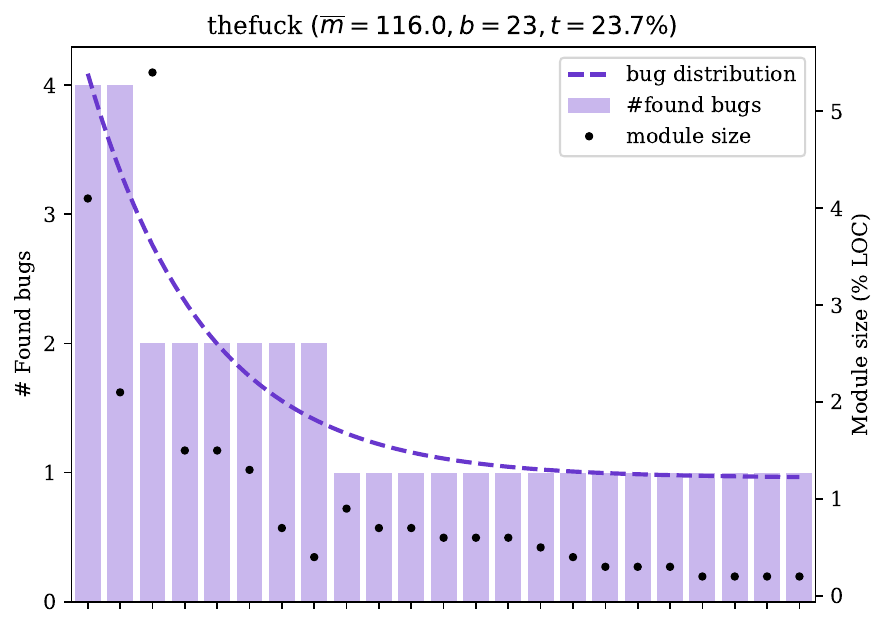}
    \end{subfigure}
    \begin{subfigure}{0.24\textwidth}
        \centering
        \includegraphics[height=3cm]{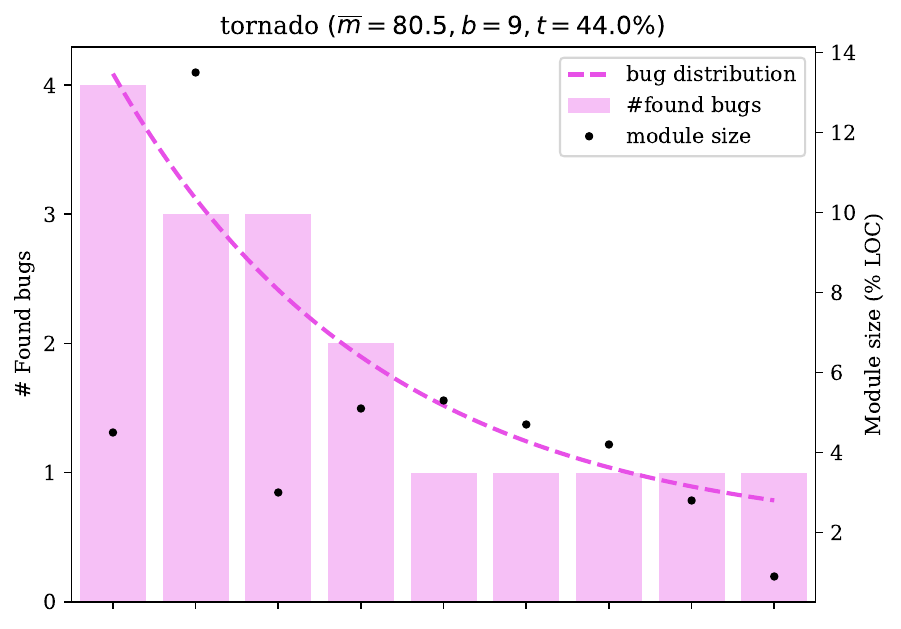}
    \end{subfigure}
    \begin{subfigure}{0.24\textwidth}
        \centering
        \includegraphics[height=3cm]{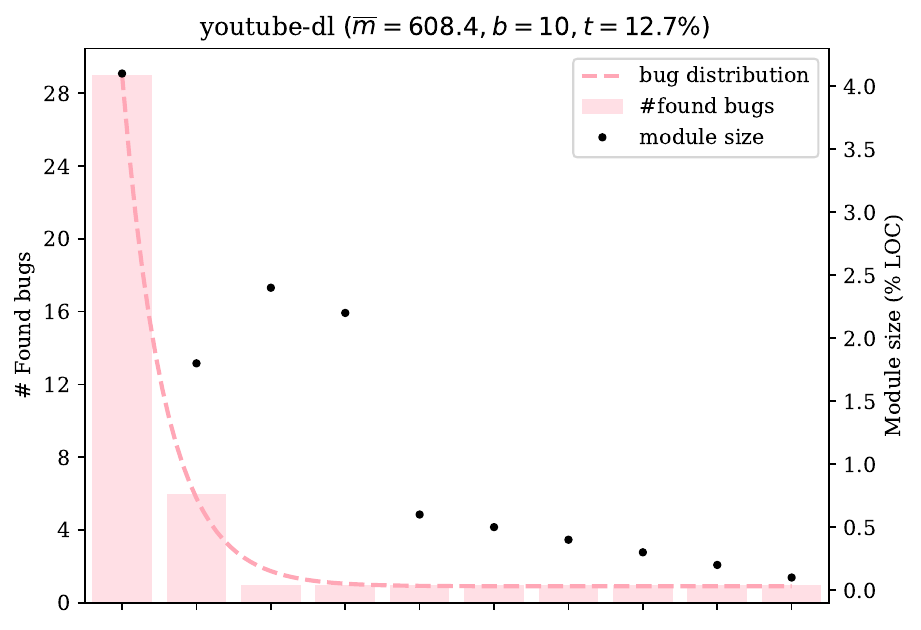}
    \end{subfigure}
    \caption{The number of identified faults in each faulty module for BugsInPy projects
      sorted by decreasing number of faults then size ($\overline{m}$: mean total
      number of modules, \textit{b}: number of faulty modules, \textit{t}: total
      faulty modules size (\% LOC))}\label{fig:bipmodule}
\end{figure*}

We turn our attention now to the spatial characteristics of the faults.
\cref{fig:d4jmodule} and \cref{fig:bipmodule} show the distribution of faults
over all modules for each project in Defects4J and BugsInPy respectively.
Comparing the number of faulty modules given in these figures with the total number
of modules per project (cf.\ \cref{tab:dataset} and the headings of
\cref{fig:d4jmodule,fig:bipmodule}), we see that the percentage of faulty modules
in each project is on average 16\% and 17\% for Defects4J and BugsInPy projects,
respectively. This result confirms the findings in current literature that a small
proportion of the code accounts for the majority of the faults~\cite{walkinshaw_are_2018,
Fenton_Ohlsson_2000, Andersson_Runeson_2007} but also shows that the fault distributions do \emph{not strictly} follow the Pareto principle, because there is no long tail: more than 80\% of the modules do not have any faults at all, rather than accounting for 20\% of the faults.

However, we note that most faulty modules from contain only a single identified fault, with the number of faults
in each module largely following a long-tailed exponential distribution. To show
this, we overlay the plots in \cref{fig:d4jmodule,fig:bipmodule} with 
lines of negative exponential best-fits. We can see in \cref{fig:d4jmodule} that most of
the Defects4J projects have a gradual slope with most modules fitting this line,
while \texttt{Time}, and several projects in BugsInPy (cf.\ 
\cref{fig:bipmodule}) have top-ranked elements far above the rest, causing
a much steeper slope that indicates the presence of \emph{bug hotspots}.

We define a \emph{bug hotspot}, in line with previous
definitions of the term \emph{hotspot}~\cite{lessler2017hotspot}, to be a module
in the system that has a higher concentration of faults than the expected number
given a random selection of events following the distribution. Since the fitted
distribution is negative exponential, any module in a project that skews this
distribution can be considered a bug hotspot. Pragmatically, we define a bug hotspot as any module that has more
than double the number of faults as modules with less faults than itself. Note
that this definition closely resembles a geometric distribution, which is the
discrete form of the exponential distribution. We define a \emph{primary} bug
hotspot to be the hotspot (if any) with the most number of faults, and a
\emph{secondary} bug hotspot to be any module that is not the primary hotspot.

Using these definitions, we see that the projects with primary bug hotspots
 are \texttt{Time} with module
\texttt{DateTimeZone} from Defects4J (\cref{fig:d4jmodule}) as well as \texttt{black} with module
\texttt{black}, \texttt{luigi} with \texttt{scheduler}, and \texttt{youtube-dl} with \texttt{utils}
from BugsInPy (\cref{fig:bipmodule}). We note that these projects are also
identifiable by their characteristically steep exponential best-fit curve. The
only project with a secondary bug hotspot is \texttt{youtube-dl} with its second most faulty
module \texttt{YoutubeDL}. We thus show that although only a small number of
modules are responsible for most of the faults in all projects, there are
relatively few bug hotspots in only a few projects.

A commonly tested and refuted hypothesis in current literature is that modules
have a higher number of faults simply because they are also larger~\cite{Fenton_Ohlsson_2000,
Andersson_Runeson_2007}, and thus their \emph{bug density} explains their number
of faults.
To investigate this hypothesis, we overlay the module sizes for each module onto
\cref{fig:d4jmodule,fig:bipmodule}. We see that for most projects this hypothesis is indeed rejected, with the 16\% of modules responsible for all faults
only contributing on average 20\% of the code. This implies a high bug density
in these modules, as is clearly evident in some modules such as the two most
faulty modules in \texttt{Math}. However, for certain projects such as \texttt{Lang}, \texttt{fastapi}, \texttt{keras},
and \texttt{pandas}, the faulty modules make up roughly 32\% of the total
number of modules but account for around 68\% of the code.
For these cases we must agree with the hypothesis, in that these modules are more
faulty due to their size.

\section{Threats to Validity}\label{sec:threats}
The main threat to the validity of this study is our exclusive use of the
multi-fault identification by An et al.~\cite{multid4j} and Callaghan and
Fischer~\cite{datasets}. Since their approaches systematically underestimate the
number of faults in each version, the results in this study are based
only on the identified faults, and may not generalize to other versions or
faults available in these projects. Our bug lifetimes are also underestimates,
and actual bug lifetimes may be larger. We ensured as far as possible that our
conclusions were based on this underestimate, to mitigate against this threat.

In \cref{sec:bug-evolution} we correlate system evolutionary changes with bug
evolutionary changes, but due to the limitations of the multi-fault
identification, this correlation may instead be due to the large system changes
causing faults to become irreproducible in these versions, while they may have
actually been introduced earlier.
To mitigate against this threat, we manually
sub-sampled faults from projects and checked if their test case
transplantation failed where fault locations continued to be identified in prior
versions. We noted that in a large number of cases, the test case transplantation
failure coincided with the fault location translation failure, i.e., the fault
became irreproducible in the version it was introduced.

The main threat to the external validity of this study is our exclusive use of
projects from Defects4J and BugsInPy. Our results may therefore
not generalize to other projects. To mitigate this threat, we compare our
results to similar studies that use different subject projects; we find that
these studies exhibit similar results and findings despite the difference in
subject projects.

\section{Conclusions}\label{sec:conclusion}
In this paper, we conducted an empirical study to understand the fault
characteristics of multiple faults in software systems contained in the popular
fault datasets Defects4J and BugsInPy.

Building on earlier work~\cite{multid4j, datasets},
we show that although Defects4J and BugsInPy \emph{identify} for evaluation purposes only
a single fault
in each entry, the underlying project versions \emph{contain} many
more faults co-existing simultaneously; the distribution indicates that multiple faults in
a version should be expected. We also find that the changes in the number of faults
in a project system are dependent on the evolution of the system, and correlate these
with the various phases of software development (feature additions, bug fixing,
etc.).

We confirm, for the projects in  Defects4J and BugsInPy, the findings of current literature
on fault lifetimes: a large number of faults are fixed within a short period
(the same day or week) after their introduction, and the remainder take much longer
to fix, up to a few months or years. However, we show that the lifetimes do \emph{not} follow a simple negative exponential distribution.

Lastly, we investigate the distribution of faults across the systems' modules. We confirm
the results of current literature on our subject projects that a minority of the
modules are responsible for the majority of the faults. However, we also find that more than 80\% of the modules contain no faults at all, thus rejecting the validity of a strict form of the Pareto principle, and that a large number of the faulty modules contain only a single fault.
However, there are a some projects in both datasets where there are particular modules are responsible for a large number of faults, indicating fault hotspots, and we give an intuitive definition of such hotspots.

Overall, our findings provide useful insight in understanding the fault characteristics
of open-source projects, particularly those used in popular fault datasets such as
Defects4J and BugsInPy. We confirm current findings on various fault
characteristics for these datasets, and provide additional insights
beyond what is found in other studies.
Our results are publicly available in our artifact~\cite{replication}.

We leave it as future work to consider other projects from the same or different
datasets, as well as to consider more versions and bugs from the underlying projects
already considered.

\bibliographystyle{IEEEtran}
\bibliography{refs}

\end{document}